\documentclass[aps,prl,reprint, superscriptaddress]{revtex4-2}
\usepackage{graphicx}
\usepackage[utf8]{inputenc}
\usepackage[T1]{fontenc}
\usepackage{xcolor, soul}
\sethlcolor{yellow}
\usepackage{soulutf8}
\usepackage{siunitx}

\begin{document}

\title{Atomic scale visualization of the \textit{p-d} hybridization in III-V semiconductors doped with transition metal impurities}

\author{K. Badiane}
\affiliation{Centre de Nanosciences et de Nanotechnologies (C2N), CNRS, Universit\'e Paris-Saclay, 10 Boulevard Thomas Gobert, 91120 Palaiseau, France}
\author{G. Rodary}
\affiliation{Centre de Nanosciences et de Nanotechnologies (C2N), CNRS,  Universit\'e Paris-Saclay, 10 Boulevard Thomas Gobert, 91120 Palaiseau, France}
\author{M. Amato}
\affiliation{Laboratoire de Physique des Solides, Universit\'e Paris-Saclay, CNRS, 91405, Orsay, France}
\author{A. Gloter}
\affiliation{Laboratoire de Physique des Solides, Universit\'e Paris-Saclay, CNRS, 91405, Orsay, France}
\author{C. David}
\affiliation{Centre de Nanosciences et de Nanotechnologies (C2N), CNRS,  Universit\'e Paris-Saclay, 10 Boulevard Thomas Gobert, 91120 Palaiseau, France}
\author{H. Aubin}
\affiliation{Centre de Nanosciences et de Nanotechnologies (C2N), CNRS,  Universit\'e Paris-Saclay, 10 Boulevard Thomas Gobert, 91120 Palaiseau, France}
\author{J.-C. Girard}
\affiliation{Centre de Nanosciences et de Nanotechnologies (C2N), CNRS,  Universit\'e Paris-Saclay, 10 Boulevard Thomas Gobert, 91120 Palaiseau, France}

\date{\today}

\begin{abstract}
$p-d$ hybridization of transition metal impurities in a semiconductor host is the mechanism that couples valence-band electrons and localized spins. We use scanning tunneling microscopy and spectroscopy combined with density functional theory to probe at the atomic scale hybridization of Cr single impurities with GaAs host. Combining spatial density of states mapping and in-gap states spectroscopy of the Cr substituted at the surface of the semiconductor, we give a detailed picture of the spatial extension and the electronic structure of the strongly anisotropic wave function of Cr on GaAs(110). First principles calculations allow to identify electronic character and origin of each states and show that the main resonance peaks and the wave function with "drop-eyes" lobes experimentally observed for $3d$ metal impurities in III-V semiconductor are direct local evidences of the $p-d$ hybridization.

\end{abstract}


\maketitle

Electronic properties of transition metal impurities introduced in a semiconductor host have been intensively studied in the 1970s and 1980s for microelectronics or optoelectronics applications (see, for instance, Refs.~\cite{Clerjaud85, Bates86}). Interest in such systems grew during the 90s because of the discovery of ferromagnetism in diluted magnetic semiconductors (DMS). The origin of magnetic order in these materials has been ascribed to carrier-mediated coupling between spins of doping transition metal dopants, which has been described by phenomenological models and first principles calculations~\cite{Dietl00, Dietl01, Mahadevan04, Sanvito01, Schulthess05}. These theoretical works have shown that the main local cause of the ferromagnetic interaction in III-V semiconductors between transition metal is the \textit{p-d} hybridization between the \textit{d} orbitals of the impurity and the \textit{p} orbitals of host atoms. 

Scanning Tunneling Microscopy (STM) and Spectroscopy (STS) contributed greatly to the understanding of the physics of transition metal ions in semiconductors by imaging at the atomic scale impurities deposited on semiconductors surfaces or substituted to cations near the surface. STM and STS studies have been carried out on the GaAs surface for several elements: Mn~\cite{Tsuruoka02, Yakunin04, Kitchen06, Jancu08, Garleff08, Richardella09, Celebi08, Taninaka16}, Fe~\cite{Richardella09, Bocquel13, Muhlenberend13}, and Co~\cite{Richardella09,Benjamin13}. Despite the different $d$-band filling, the STS data show similar resonances in the band-gap and the conductance maps at the corresponding resonance voltages reveal extended wave-functions with "crab-like" and "drop-eye" shapes. These results have been discussed qualitatively following the Mahadevan and Zunger approach developed for bulk materials~\cite{Mahadevan04}, but no precise explanation of the origin of the shapes and their relations to the contributions of the $p$ and $d$ orbitals to the state has been so far given for surfaces. In spite of the large number of STM/STS studied showing similar results on transition metal impurities in III-V semiconductors surface, a complete description of the physical mechanism explaining the observed wave function shape and resonance states is still missing. In this letter, our analysis of the extended impurity states present an intermediate description between the local chemical description, i.e. the $d$-orbitals of the impurity hybridize with the $p$-orbitals of the neighboring anions, and the physical description, i.e. the $d$-orbitals hybridize with Bloch valence states at the $\Gamma$ point. Such a description will enable a more precise understanding of the kinetic exchange mechanism enabling ferromagnetic ordering in those materials.

We explore by STM and STS the electronic states of Cr single atoms substituted to Ga in the first layers of GaAs(110) surface and we combine experimental results with Density Functional Theory (DFT) calculations. To our knowledge, Cr atoms have never been explored by STM on GaAs. It has been theoretically demonstrated that they behave as Mn and give rise to ferromagnetic coupling~\cite{Mahadevan04} but only few DFT studies focus on the electronic properties of transition metal impurities near a semiconductor surface \cite{Fang16Fe, Fang16Co, Islam2012}. Experimentally, while Cr and Mn induce different modifications in the electronic structure (energy position of the acceptor level, order of the resonance states), they both show effective magnetic coupling at low temperature~\cite{Wu11}. Thus, our study of Cr by STM/STS allows to complete the understanding of the mechanism behind the transition metal doping of semiconductors. 

To control the substitution of Ga by Cr in the first and second layers of the surface, we developed a tip-induced manipulation method. This gives the possibility of studying magnetic impurities as function of their coordination number: three covalent bonds with the As neighbors in the case of Cr inserted in the first layer, noted $Cr_{Ga1}$ and shown in Fig. 1(a), and four bonds in the case of Cr in the second layer, noted $Cr_{Ga2}$ and shown in Fig. 1(b), which has the same coordination number than in the bulk. For these two configurations, we image the wave function of the impurity and measure spectra. Combining the experimental spatial and energy information with first principles total energy calculations, we obtain spatial maps of the different contributions, $s$, $p$, $d$ of the wave function, which provides a detailed view of the hybridization between the $d$-orbital of the Cr atom and the $s-p$ orbitals of the anion.

Experiments were performed with a STM system at $4.2$~K in ultrahigh vacuum ($<5\times10^{-11}$ mbar). The gap voltage was applied on the sample. We have cleaved \textit{in situ} p-type highly doped ($2\times10^{19}$ cm$^{-3}$) GaAs wafers in order to expose the (110) surface and to perform so-called cross-sectional STM (X-STM)~\cite{Girard09}. The sample was pre-cooled  in the STM cryostat and then transferred to a preparation chamber where a submonolayer quantity of Cr was deposited on the GaAs(110) surface, and replaced quickly in the STM for measurement at $4.2$ K. As described by D. Kitchen et al.~\cite{KitchenThesis06, Kitchen06, Richardella09}, transition metal adatoms deposited on GaAs(110) can be incorporated in the first surface layer. Mn, Fe and Co adatoms on GaAs(110) show different surface dynamic and incorporation behavior. For Cr, we find that adatoms can be incorporated in the first GaAs layer using tip manipulation (Fig. 1(a)). Cr in the first layer can also be forced to move to substitution position in the second layer (Fig. 1(b)). In each case of substitution, we observe the extraction of a Ga atom visible as an adatom: black rings also reported in \cite{Kitchen06} and \cite{Richardella09}. Mechanism of tip-induced substitution and atoms manipulation will be described elsewhere~\cite{Rodary}.

\begin{figure}
\includegraphics[width=70mm]{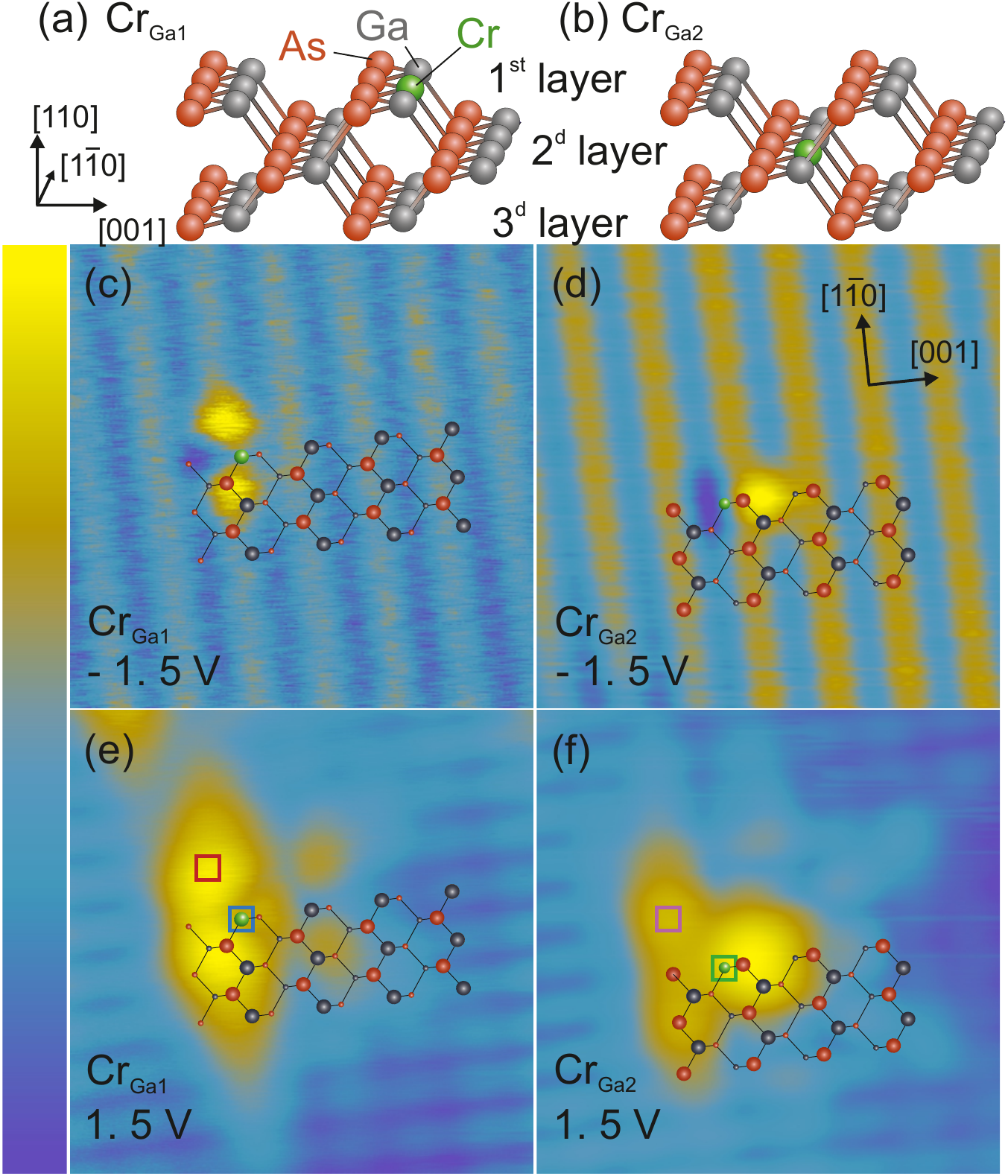}
\caption{\label{Figure 1}(a) and (b): Schematic 3-dimensional view of the first three layers of the GaAs(110) surface with one Cr atom incorporated. As atoms are in red, Ga in grey and Cr in green. (c)-(f): Constant current STM image of one Cr atom substituted to a Ga in the first layer of GaAs(110), (c) and (e), and in the second layer, (d) and (f). Ball and stick atomic models are superposed to the STM images; large balls correspond to atoms in the first layer while smaller balls correspond to atoms in the second layer as defined in (a) and (b). Parameters: 3.5~nm$ \times$ 3.5~nm, (c) $-1.5$~V, $50$~pA, z range $39$~pm, (d) $-1.5$~V, $100$~pA, z range $60$~pm, (e) $1.5$~V, $50$~pA, z range $360$~pm, (f) $1.5$~V, $60$~pA, z range $444$~pm}
\end{figure}

Figure 1 (c-f) shows constant current STM images of Cr atoms in Ga substitution position in the first and second layers of GaAs(110) for negative and positive tip-sample gap voltage. Because we control the incorporation with the tip, we are able to determine the exact Cr position before and after the incorporation processes. As expected for Ga substitution, Cr is localized in the Ga [1-10] columns, which are imaged at positive voltage \cite{Feenstra87}. $Cr_{Ga1}$ is imaged on a Ga [001] row of the first layer (Fig. 1 (e)) while $Cr_{Ga2}$ appears in between two Ga rows, i.e. in a As row of the first layer and Ga row of the second layer (Fig. 1 (f)). By symmetry with respect to these atoms rows, we can deduce that the Cr is inserted in the second layer. In the first layer, we find similar patterns with drop-eyes (negative voltage) and crab-like (positive voltage) shape as found on other transition metal impurities \cite{Garleff08,Richardella09, Muhlenberend13, Benjamin13}; in the second layer the patterns change drastically. Such observation of a transition metal atom in the second layer has been reported for Mn in \cite{Garleff08}, but no atomic resolution and no bias voltage dependence have been shown. It is noteworthy that in both layers the maximum of electronic density of states is not localized exactly on the Cr atoms but  distributed rather asymmetrically over several atomic length in a strong anisotropic way.

\begin{figure}
\includegraphics[width=70mm]{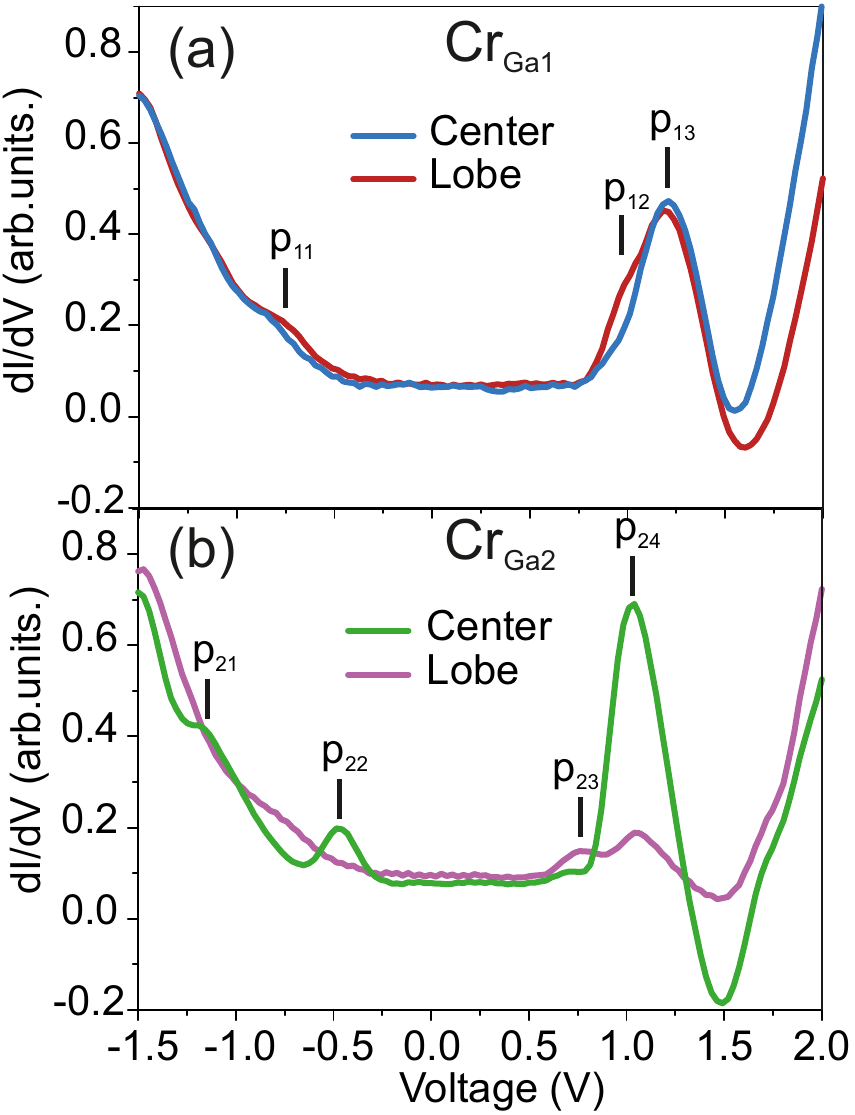}
\caption{\label{Figure 2}STS spectra measured on top of Cr atoms incorporated in the first layer of GaAs(110) surface (a) and in the second layer (b). Exact position of the tip during the spectra, named Center or Lobe, are indicated by small squares depicted in Fig. 1(e) and (f). Peaks observed on spectra are indicated and labeled on each peak.}
\end{figure}

 \begin{figure}[t]
\includegraphics[width=70mm]{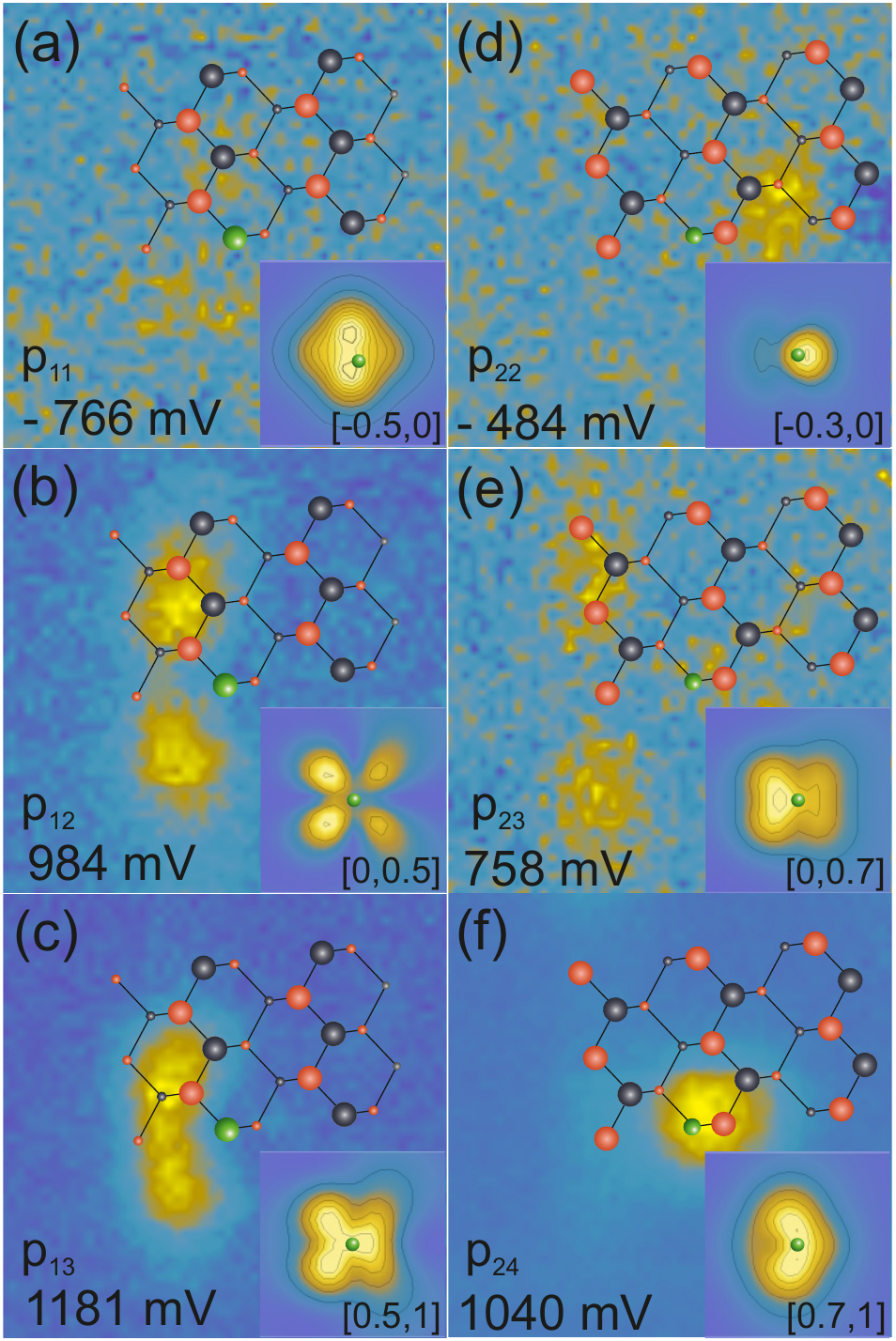}
\caption{\label{Figure 3}$dI(V)/dV$ maps of Cr in Ga substitution position in the first surface layer (top row) and in the second layer (bottom row), 2~nm$ \times$ 2~nm. Images are measured at the energies of the peaks observed in Fig. 2. Schematic ball and stick atomic model are superposed to STM images. Insets of each $dI(V)/dV$ map: calculated cross-section on the $(110)$ plan of ILDOS of Cr in the first surface layer (left) and in the second surface layer (right) at energy ranges corresponding to each peaks of Fig. 4, indicated in brackets. Maps are integrated over a slice along $[110]$ from the surface to the plan of the $Cr_{Ga2}$ atom. Isodensity contours are marked by black lines and are separated by $0.005$ $e/bohr^{3}$. Green dots indicate the position of Cr atom. The whole color range on each ILDOS map is (in $e/bohr^{3}$ unit): $0.054$ in (a), $0.011$ in (b), $0.029$ in (c, $0.026$ in (d), $0.028$ in (e), $0.031$ in (f).}
\end{figure}

Figure 2 presents differential conductance $dI(V)/dV$ spectra measured on Cr, proportional to the local electronic density of states (LDOS). They show features similar to those measured on Mn, Fe, and Co ~\cite{KitchenThesis06, Richardella09, Muhlenberend13, Benjamin13}. Negative differential conductance is also observed due to the presence of a narrow resonance peak \cite{Xue1999}. Depending on the exact spectra position, center or lobe, large in-gap resonance are observed in the GaAs gap near the conduction band and smaller peaks in the valence band.

In order to precisely identify the spatial distribution of each state, we performed differential conductance maps on a $52 \times 52$ grid. A conductance map taken at the energy of an electronic resonance represents the probability density of the corresponding wave function. Figure~3 shows $dI(V)/dV$ maps of $Cr_{Ga1}$ (left) and $Cr_{Ga2}$ (right) at the energies of the peaks observed in Fig. 2 (except peak $p_{21}$ for which the signal is too low). For $Cr_{Ga1}$, the first state $p_{11}$ (Fig.3(a)) is weak and delocalized. In contrast, the second state $p_{12}$ (Fig.3(b)) is localized near the As neighbour atoms. The $p_{13}$ state (Fig. 3(c)) is similar to $p_{12}$ but present in addition a signal contribution in between the As atoms, closer to the Cr position. For $Cr_{Ga2}$, the states $p_{22}$ (Fig. 3(d)) and $p_{24}$ (Fig. 3(f)) present an isotropic spatial extension while $p_{23}$ (Fig. 3(e)) has an delocalized wave function. The anisotropic shape of the Cr atom in the constant current image of Fig. 1(f) can be well reproduced by the summation of the conductance images at positive bias of Fig. 3(e) and 3(f).

To further explain and clarify the origin of the observed peaks in Fig. 2 and their spatial distribution in Fig. 3, we have carried out first principles total energy calculations in the framework of Density Functional Theory (DFT) as implemented in the the plane-wave PWscf code of the QuantumESPRESSO distribution~\cite{PWSCF2009}. Local Spin Density Approximation (LSDA) with norm-conserving pseudopotentials with the Perdew-Zunger exchange-correlation functional were employed. We simulated the effect of Cr dopants in both bulk and (110) GaAs surface. For each system the Projected Density of States (PDOS) was calculated by projecting the Kohn-Sham wave functions on an atomic orbitals basis set in order to evaluate the contribution of each atom to the total DOS. Experimental constant-current STM images are here discussed in light of Integrated Local Density of States (ILDOS) calculations in the framework of the Tersoff-Hamann approximation ~\cite{TersoffPRB1985, GiustinoOxford2014}. According to this assumption, the STM tunneling current is proportional to the integrated LDOS of the sample at the tip position. By employing this approach, constant-current STM images can  hence be directly compared to the integrated LDOS calculated within DFT. All the computational details of our study can be found in the Supplemental Material (SM). 

 \begin{figure}
\includegraphics[width=70mm]{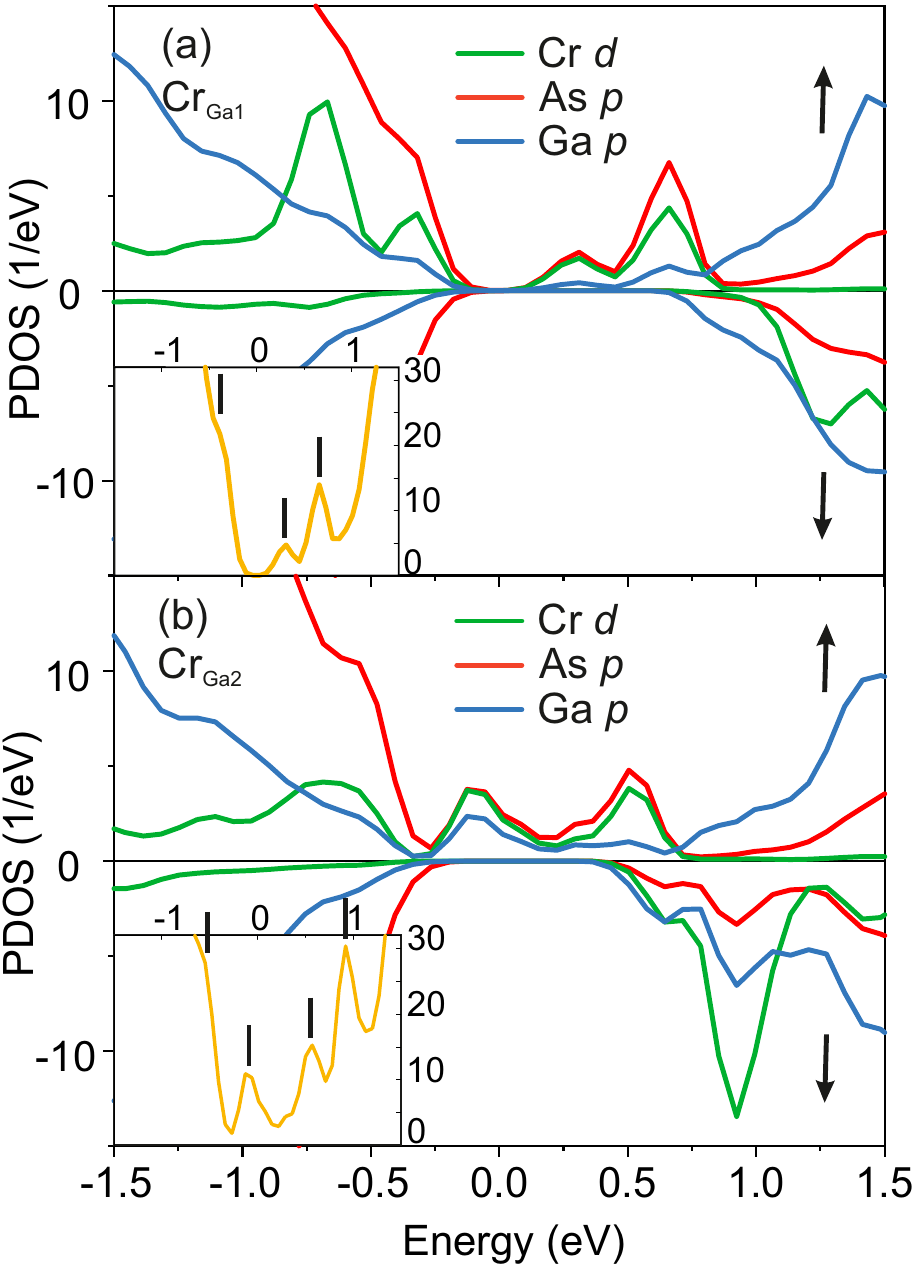}
\caption{\label{Figure 3} Calculated spin resolved density of states projected on the atomic orbitals for the Cr-doped (110) GaAs surface with the Cr atom in the first (a) and second (b) layer of the surface. The insets show the total density of state and vertical black bars show the position of the maxima. The zero of the energy is set to the Fermi energy of the system. Spin up correspond to upper curves, and spin down to the lower ones.}
\end{figure}

  We calculated the electronic density of states projected on atomic orbitals for Cr-doped bulk GaAs and for a Cr atom in the first, second and third layer of the GaAs surface. In the bulk case (see SM), the Cr site has the $T_{d}$ symmetry and produces a deep trap state within the gap that results from the hybridization of the Cr $d$ orbitals and the $p$ orbitals of the As atoms (see Fig. S1 and Table S1 of the SM). The tetrahedral crystal field of the zinc-blend host splits the $d$ levels of the Cr atoms into $e$ and $t_{2}$ multiplets, which are further split by the exchange interaction into spin up and spin down states. Hybridization finally splits the $t_{2}$ states into bonding and antibonding levels ($e$ states have no orbital in the GaAs of the same symmetry to hybridize with), giving rise to six states in total (Figure S2 and S7 of the SM). According to \cite{Mahadevan04}, this leads to two kind of impurity states: the crystal field resonances (CFR) that have $d$-orbital character and are localized on the transition metal ion and the dangling bond hybrid (DBH) that contains a large $p$-orbital component due to the $p-d$ hybridization.
  
  When the Cr atom substitutes a Ga atom in the third layer of the surface (Fig.~S6), the electronic structure does not differ too much from what we observe in the case of Cr-doping of the bulk system. As one goes from the third to the second and first layer of the surface, the symmetry of the Cr site is lowered as evidenced by the evolution of first neighbours distances around the Cr impurity (see Table S2). For the $Cr_{Ga1}$ substitution in the first layer, the three-fold coordination and the surface buckling leads to $C3v$ site-symmetry. Furthermore, because of the change of the lattice parameters at the surface, a $C_{S}$ site-symmetry is finally obtained. As a consequence, the five-fold degeneracy of the $d$-orbital multiplet is completed lifted as shown in Fig. S7.

 Insets of Fig. 4 show the total density of states for the $Cr_{Ga1}$ and $Cr_{Ga2}$ that exhibits three peaks for $Cr_{Ga1}$ and four for $Cr_{Ga2}$, corresponding to what is observed experimentally in Fig.~2. Figure~4 shows the contributions of atomic orbitals to the total density of state ($d$ and $p$ orbitals, while $s$ orbitals have a negligible contribution), for the two spin polarizations. All the peaks observed have a strong $d$ contribution coming from the chromium, as well as a As-$p$ contribution. For $Cr_{Ga1}$ (respectively $Cr_{Ga2}$), this $p$ contribution is especially strong for the peaks at $-0.3$ eV ($-0.6$ eV) and $0.7$ eV ($0.5$ eV). For the peak at $0.7$ eV for $Cr_{Ga1}$ and the peaks at $-0.1$ eV and $0.8$ eV for $Cr_{Ga2}$, the contribution of Ga-$p$ orbitals is also significant as compared to the $d$ contribution. This mixed character of the peaks is the consequence of $p-d$ hybridization.
 
 Insets of Figure 3 shows ILDOS of $Cr_{Ga1}$ (left) and $Cr_{Ga2}$ (right) at energies corresponding to the peaks of Fig. 4 and associated to experimental maps. Here, we represent $(110)$ plans of ILDOS integrated over a slice along $[110]$ from the surface to the plan of the $Cr_{Ga2}$ atom. These calculations reproduce well the $dI/dV(V)$ experimental images, considering the well-known DFT inaccuracy in reproducing experimental energy peak positions, except for in Fig. 3 (a) where the signal is too weak. For $Cr_{Ga1}$, the observed and simulated lobes are oriented towards the As first neighbours of the Cr, which can be explained by the strong As-$p$ contribution to the density of state found in Fig. 4. Thus, the DFT results show that our STM observation of the delocalization and the anisotropy of the wave function, especially the density of states oriented towards the As neighbors giving rise to the two drop-eyes lobes, is a direct demonstration of $p-d$ hybridization observed here at the atomic scale. 
 
Some states present wave functions with a rather compact shape centered around the Cr (insets of Fig. 3 (a), (d) and (f)), while others are more delocalized with anisotropic lobes (insets of Fig. 3 (b), (c) and (e)). States centered on the Cr impurities have a CFR character, with a rather out-of-plane contribution, which could be described by $d_{z^{2}}$ orbitals. In contrast, states with lobes pointing towards the anions have a DBH character with the symmetry of $d_{xy}$ orbitals. One expect in this case a naturally strong sigma bonding with the $p$ orbitals of the As neighbors which have the same symmetry.

Because of the lowering of site-symmetry from $T_{d}$ in the bulk to $C_{S}$ at the surface, the energy levels identified by STM spectroscopy and reproduced by the DFT calculations cannot be classified according to the irreducible representations $t_{2}$ and $e$ of the $T_{d}$ point-group. Despite this, one can still identify two types of states, the localized states, essentially of $d$-orbital character, and the extended hybridized $p-d$ states. Because these extended states spread over several unit cells, they should be better described as hybridization of $d$-orbitals with valence Bloch states at the $\Gamma$ point \cite{Koiller2001,Voisin2020}.

In conclusion, by combining STM and first principle calculation, we were able to explain the origin of the anisotropic shape of the wave function of the Cr atom substituted to Ga in GaAs(110) surface. The lobes observed by STM in GaAs doped by transition metals are the signature of the electronic hybridization of the atom with its host. In particular, the As-$p$ orbitals hybridize with the Cr-$d$ orbitals giving rise to "drops-eyes" lobes localized around As atoms, which are the main distinctive attribute of the wave function of transition metals impurities in GaAs. Our DFT calculations describe the electronic origin of the resonances observed in the semiconductor gap when doped with a transition metal, and show the significant weight of the As-$p$, as well as the Ga-$p$ electrons in some case. We also present the spin resolved contribution of these states, whose experimental confirmation could be an interesting perspective of this work.

\begin{acknowledgments}
The work was supported by the French ANR project MechaSpin (ANR-17-CE24-0024). The high-performance computing (HPC) resources for this project were granted by the Institut du developpement et des ressources en informatique scientifique (IDRIS) under the allocations A0040907723 and c2016097723 via GENCI (Grand Equipement National de Calcul Intensif).
\end{acknowledgments}

\bibliographystyle{apsrev4-2}
\bibliography{manuscript}

\end{document}